\documentclass{article}

\usepackage{graphicx}

\newlength{\mywdithgraph}
\begin{document}

\title{Effects of Metallic Gates on ac Measurements of the Quantum 
Hall Resistance} 
\author{Fr\'ed\'eric~Overney,~Blaise~Jeanneret~and~Beat~Jeckelmann
\thanks{The authors are with the Swiss 
Federal Office of Metrology and Accreditation, Lindenweg 50, CH-3003 
Bern-Wabern, Switzerland (\mbox{e-mail:} frederic.overney@metas.ch).}} 

\maketitle
%----------------------------------------------------------------------
% Abstract:
%---------------------------------------------------------------------- 
\begin{abstract}
Using a sample with a split back-gate, a linear frequency dependence of the
ac quantum Hall resistance was observed. 
The frequency coefficient, which is due to dielectric losses
produced by leakage current between the 2DEG and the back-gates, can
be turned from a positive to a negative values by increasing the 
back-gate voltage.  More interestingly, by removing theses back-gates, the
losses can be considerably reduced leading to a residual frequency
coefficient on the order of (0.03$\pm$0.03)$\cdot$10$^{ -6}$/kHz. 
Moreover, at 1 kHz, an extremely flat plateau was observed over a magnetic field 
range of 1.4 T. These results clearly indicate that the audio frequency dependence
of the QHR is to a large extend related to the measurement apparatus and
does not originate from the physical transport properties of the 2DEG.
\end{abstract}

% \begin{keywords}
% Coaxial ac bridge, metalic gates, quantum Hall effect, dielectric losses, 
% frequency dependence. 	
% \end{keywords}

%----------------------------------------------------------------------
% SECTION: Introduction
%---------------------------------------------------------------------- 
\section{Introduction}

The universal nature of dc electron transport in a two dimensional
electron gas~(2DEG) at low temperature and high magnetic field makes
the quantum Hall resistance (QHR) an ideal primary standard of
resistance (see \cite{Jeckelmann01b} for the latest review on the
application of the quantum Hall effect in metrology).

The situation in the regime of ac transport is rather different.  The
pioneering work of Melcher \cite{Melcher93b} showed that the QHR
measured at a frequency of $1592~\rm{Hz}$ agrees with $R_{\rm{K}}/2$
($R_{\rm{K}}\equiv h/e^{2}$ is the von Klitzing constant) with an
relative standard uncertainty of $3~\mu \Omega /\Omega$.  This
work has triggered a series of investigations at several National
Metrology Institutes.  Although various controversial results have
been obtained, a few common features were observed
\cite{Jeckelmann01b} : the plateaus in the quantum Hall
resistance $R_{\rm{H}}$ are no longer as flat and broad as they are at
dc.  In addition, the QHR has a linear frequency dependence caused by
losses along the sample edges due to the presence of metallic gates in
the vicinity of the 2DEG \cite{Delahaye00c,Schurr01}.

In this paper, we show that, by removing metallic gates from the
vicinity of the sample, the frequency dependence of the QHR can be
strongly reduced without any external potential adjustment.  Under
appropriate conditions, the losses can be reduced to a level where the
ac QHR can be used as an ac resistance standard with an uncertainty of
a few parts in $10^{8}$ at kHz frequencies over a magnetic field
range broader than 1 T. These results suggest that the frequency
dependence observed in earlier ac measurements of the QHR originate to a large
extend in the measurement apparatus and not in the physical properties of
the 2DEG.

%----------------------------------------------------------------------
% SECTION: Modelisation of Losses
%---------------------------------------------------------------------- 
\section{Model for ac Losses}
\begin{figure}
\includegraphics[width=\mywdithgraph]{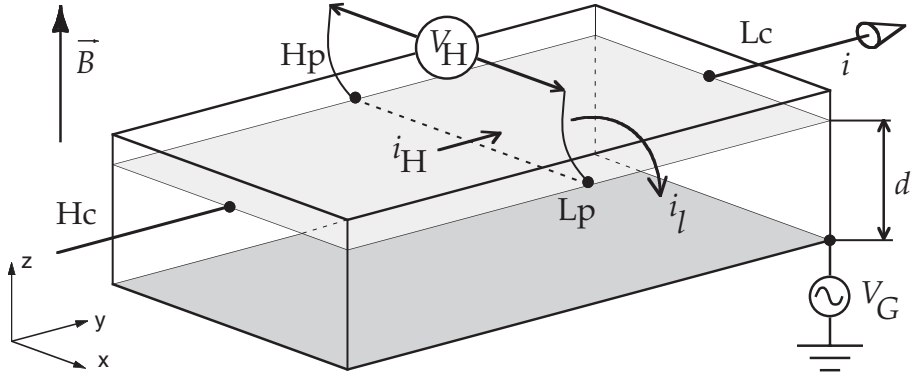}
\caption{Schematic of a AlGaAs/GaAs heterostructure showing
the loss mechanism due to the capacitive current flowing from the
2DEG to the back-gate.}
\label{LossModel}
\end{figure}

In figure~\ref{LossModel}, a two-dimensional electron gas~(2DEG),
located in a AlGaAs/GaAs heterostructure is sketched.
The current, directed along the y-axis, enters the 2DEG
through the contact $\rm{H_{c}}$ and leaves it through the contact
$\rm{L_{c}}$.  For a magnetic field directed along the z-axis, the low
and high potential contacts are $\rm{H_{p}}$ and $\rm{L_{p}}$ respectively. 
A back-gate is located at a distance $d$ underneath the 2DEG and kept 
at a potential $V_{\rm{G}}$.
The Hall impedance $Z_{\rm{H}}$ of the device is defined by the ratio
of the Hall potential $V_{\rm{H}}$, measured between the potential
contacts $\rm{H_{p}}$ and $\rm{L_{p}}$, to the current $i$ leaving the
device at the current contact $\rm{L_{c}}$.  In this model, the
deviation of $Z_{\rm{H}}$ from the Hall resistance $R_{\rm{H}}$ is due
to leakage current.  More precisely, the current $i$ leaving the
device at $\rm{L_{c}}$ is not equal to the Hall current $i_{\rm{H}}$
generating the Hall voltage because a small capacitive leakage current
$i_{l}$ flows to the backgate and does not contribute to $i$. 
Therefore, considering that $V_{\rm{H}}=R_{\rm{H}} i_{\rm{H}}$:
\begin{equation} \label{za}
	Z_{\rm{H}} = \frac{V_{\rm{H}}}{i} = \frac{V_{\rm{H}}}{i_{\rm{H}}-i_{l}}\approx 
	R_{\rm{H}}(1+\frac{R_{\rm{H}}}{V_{\rm{H}}} i_{l})
	=R_{\rm{H}}(1+\Delta)
\end{equation}
Considering that each elementary surface, $dxdy$, of the 2DEG forms a
parallel plate capacitor with the back-gate, the leakage
current $i_{l}$ can be expressed as
\begin{eqnarray} \label{il}
	i_{l}&=& j \omega \tilde{\epsilon} V_{\rm{H}}
	\frac{\epsilon_{o}}{d} 
	\int\!\!\!\int_{S}(\frac{V(x,y)-V_{G}}{V_{\rm{H}}})dxdy \\
	&=& j \omega \tilde{\epsilon} C_{o}(B,V_{G})V_{\rm{H}}
\end{eqnarray}
where $j=\sqrt{-1}$, $\omega=2 \pi \nu$ is the angular frequency,
$\tilde{\epsilon}$ is the complex dielectric constant of the
heterostructure, $\epsilon_{o}$ is the permittivity of free space and
$V(x,y)$ is the local potential of the 2DEG. The term
$C_{o}(B,V_{\rm{G}})$ can be interpreted as the equivalent free space
capacitance between the 2DEG and the backgate.  It is not a purely
geometrical capacitance but rather an electrochemical capacitance in
the sense of \cite{Christen96b}.  Therefore, if the potential of the
backgate is sufficiently high, the sign of the current is inverted and
an extra current is now injected in the 2DEG from the backgate. 
Accordingly, $C_{o}(B,V_{\rm{G}})$ is then negative.

The integration surface $S$ extends from the line between the
potential contacts, $\rm{H_{p}}$ and $\rm{L_{p}}$ and the sample edge
where the current leaves the 2DEG through the $\rm{L_{c}}$ contact. 
Any current flowing between the 2DEG and the backgate before the
potential contact does not induce any error.  Indeed, if a leakage
current leaves the 2DEG before the potential contact, it will neither
generate a Hall voltage nor contribute to the current $i$.  Similarly,
if a current is injected into the 2DEG from the backgate before the
potential contacts, it will generate an Hall voltage and also
contribute to the current $i$.

The leakage current $i_{l}$ is mostly in quadrature with the Hall
current $i_{\rm{H}}$, however, dielectric losses in
the GaAs heterostructure will generate a small in-phase component. 
Writing the complex dielectric constant as
$\tilde{\epsilon}=\epsilon(1-j \tan(\delta))$, $\epsilon$ being the
real component of the dielectric constant and $\tan(\delta)$ the dielectric 
losses, the
correction term $\Delta$ in (\ref{za}) becomes
\begin{equation} \label{Delta}
	\Delta = \omega R_{\rm{H}}\epsilon \tan(\delta) C_{o}(B,V_{G})+j
	\omega R_{\rm{H}}\epsilon C_{o}(B,V_{G})
\end{equation}
The loss mechanisms introduce a linear frequency dependence in $\Re
e\{Z_{\rm{H}}\}$ that can be reduced by making $C_{o}(B,V_{\rm{G}})$
as small as possible.

%----------------------------------------------------------------------
% SECTION: The ac Bridge and the connection scheme
%---------------------------------------------------------------------- 
\section{The ac bridge and the connection scheme}
The ac quantum Hall effect was investigated by measuring the ratio
between $Z_{\rm{H}}$ measured on the plateau $i = 2$ and a quadrifilar
resistor $Z_{\rm{G}}$ of the same nominal value having a calculable ac/dc
difference \cite{Gibbings63}. 

\begin{figure}%[h]
\vspace{-1em}
\includegraphics[width=\mywdithgraph]{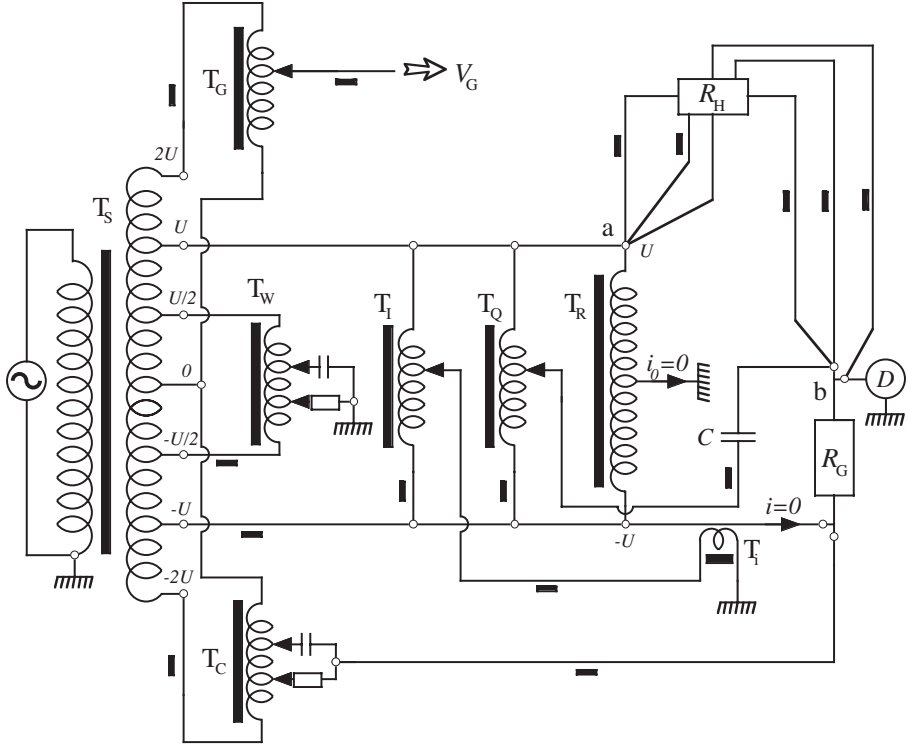}
\caption{Schematic of the ac bridge.}
\label{acBridge}
\end{figure}

Figure~\ref{acBridge} shows the schematic of the bridge constructed
according to the usual ac coaxial bridge techniques \cite{Kibble84}. 
For simplicity, the outer coaxial conductor is omitted.  The bridge is
powered by the double screened transformer $\rm{T_{S}}$ featuring taps
with output voltages equal to $\pm 2U$, $\pm U$, $\pm U/2$ and $0$,
where $U$ is the nominal voltage applied to the QHR and to
$R_{\rm{G}}$.  The Wagner autotransformer $\rm{T_{W}}$ is used to
compensate the leakage current until the load current $i_{0}$ at the
centre tape of the ratio autotransformer $\rm{T_{R}}$ is zero.  The
combining network $\rm{T_{C}}$ acts as a current source and is
adjusted to zero the current $i$ through the lead between the high
potential port of $R_{\rm{G}}$ and the $-U$ tap of $\rm{T_{R}}$.  The
ratio of the resistors has a small deviation from unity that is
compensated using the autotransformer $\rm{T_{I}}$ and the injection
transformer $\rm{T_{i}}$ for the in-phase component, and using the
autotransformer $\rm{T_{Q}}$ and the low loss capacitor $C$ for the
quadrature component.  The offset of the 1:1 ratio of the
autotransformer $\rm{T_{R}}$ is measured by repeating the comparison with
$\rm{T_{R}}$ reversed.  Finally, the autotransformer $\rm{T_{G}}$ sets the
potential $V_{\rm{G}}$ of the back gate when gated samples are measured.
The quantum Hall sample is connected to the bridge using a multiple
series connection \cite{Delahaye93b}.  In such a connection scheme,
the current and potential leads of the same polarity are tied together
at the external junction points a and b where the apparent quantum
Hall impedance $Z_{\rm{H}}$ can be defined as a two terminal-pair
component $Z'_{\rm{H}}$.  The quadrifilar resistor $Z_{\rm{G}}$ is defined
as a four terminal-pair component $Z'_{\rm{G}}$.  Accounting for cable
correction one obtains \cite{Chua99,Schurr01}
\begin{eqnarray} \label{Zm}
	Z'_{\rm{H}} &=& 	
	Z_{\rm{H}}\cdot(1+\frac{Y_{\rm{H_{p}}}Z_{\rm{H_{p}}}}{2})
	\cdot(1+\frac{Y_{\rm{L_{c}}}Z_{\rm{L_{c}}}}{2})
	\nonumber\\
	& & {}\cdot(1+(\frac{Z_{\rm{H_{p}}}}{Z_{\rm{H}}})^n+
	(\frac{Z_{\rm{L_{c}}}}{Z_{\rm{H}}})^m)\\
	Z'_{\rm{G}} &=& 
		Z_{\rm{G}}\cdot(1+\frac{Y_{\rm{H_{p}}}Z_{\rm{H_{p}}}}{2})^{-1}\cdot
		(1+\frac{Y_{\rm{L_{c}}}Z_{\rm{L_{c}}}}{2})^{-1}
\end{eqnarray}
where the terms $YZ=\omega^2 L C + j \omega R C$ are related to the
leads which define the high potential (index $\rm{H_{p}}$) and the low
current (index $\rm{L_{c}}$).  The cable corrections are dominated by the
terms involving the long coaxial cables linking the quantum Hall
device, inside the cryostat, to the junction points a and b
outside the cryostat.  The real component of the total cable correction is
proportional to the square of the frequency and amounts to $9.2\cdot
10^{-8}/\rm{kHz^2}$.

The exponents $n$ and $m$ in (\ref{Zm}) are related to the multiple
series connection scheme and denote the number of leads connected to
the high- and low-junction points (a and b respectively). 

%----------------------------------------------------------------------
% SECTION: Experimental Results
%---------------------------------------------------------------------- 
\section{Experimental Results and Discussions}
Measurements have been carried out on three GaAs heterostructures
(Type LEP~514 \cite{Piquemal93}).  The first device, LEP1, is mounted
on a printed circuit board equipped with a classical back-gate
connected to the shield.  The second device, LEP2, is mounted on a
printed circuit board equipped with a split back-gate in a way similar
to \cite{Delahaye01}.  While the back-gate underneath the low
potential side of the QHR was always grounded, the potential of the
back-gate below the high potential side was set to a potential $V_{\rm{G}}$
ranging between $0$ and $2 V_{\rm{H}}$ using the autotransformer
$\rm{T_{G}}$.  The third device, LEP3, is mounted on a printed circuit
board where the back-gate was removed and all metallic parts were kept
as far as possible from the sample.  At a temperature of $0.3~\rm{K}$,
the frequency dependence of the Hall impedance $Z_{\rm{H}}$ on the plateau
$i=2$ was investigated with an ac current of $20~\mu \rm{A}$.  The
frequency was varied between 800~Hz and 5~kHz.

\subsection{Changing the Position of the Potential Contact}
\begin{figure}%[!b]
\includegraphics[width=\mywdithgraph]{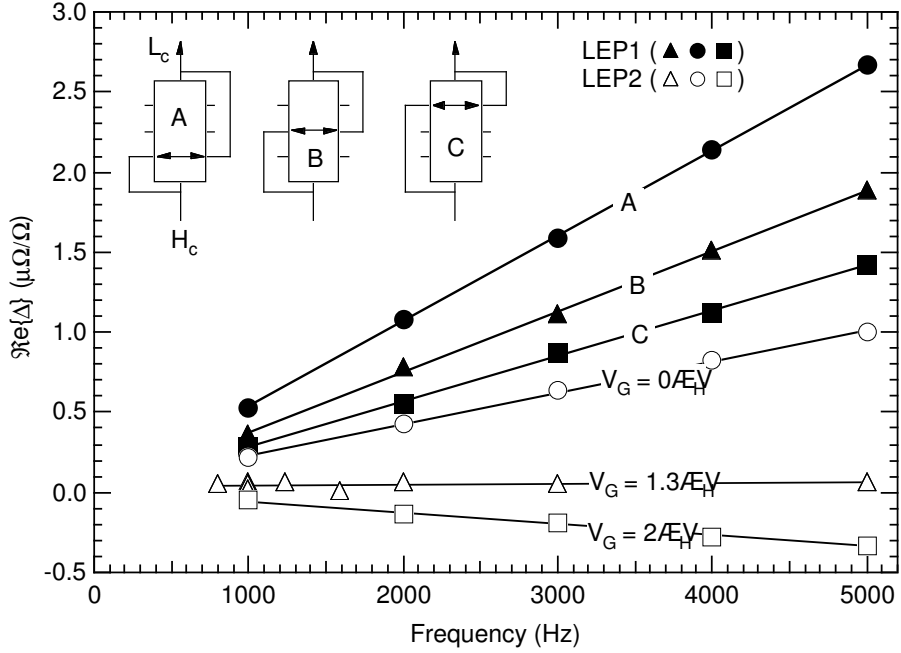}
\caption{Relative deviation of the real component of $Z_{\rm{H}}$,
$R_{\rm{H}}(1+\Re e\{\Delta \})$, from its extrapolated DC value,
$R_{\rm{H}}$, measured at 20~$\mu$A on the centre of plateau $i$=2 of
two different samples.  Solid symbols: LEP1 using the three different
double-series connections showed in the inset (the magnetic field
points out of the page).  Open symbols: LEP2 applying three
different voltages to the back-gate, the sample being connected with
the triple series-connection.}
\label{RHvsFreq}
%\vspace{-0.3cm}
\end{figure}

The frequency dependence of the Hall impedance of the sample LEP1 in
three different double-series connections is shown in
Fig.\ref{RHvsFreq} (solid symbols) for three different pairs of potential contacts.

A linear frequency dependence is observed in the real component of
$Z_{\rm{H}}$.  For the configuration A the linear frequency
coefficient $\alpha \equiv \partial \Re e\{Z_{\rm{H}}\}/ \partial \nu
$ amounts to 0.53~$\cdot 10^{-6}/$kHz and decreases to 0.28~$\cdot
10^{-6}/$kHz for the configuration C. The linear frequency coefficient
$\beta \equiv \partial \Im m\{Z_{\rm{H}}\}/ \partial \nu$ shows a
similar behaviour.

As expected from our model, the total leakage current $i_{l}$
decreases when the potential contacts are moved toward $\rm{L_{c}}$, i.e.
the integration surface $S$ decreases, therefore reducing the linear
frequency coefficient, $\alpha$ and $\beta$.  The variation of $\alpha$
(respectively $\beta$) as a function of the distance between the
potential contacts and $L_{c}$ is close to linear.  This means that the
equivalent free space capacitance $C_{o}$ is quite homogneously
distributed along the sample at least in the central part of the
device, i.e. far from the current contacts.  According to
(\ref{Delta}), the ratio of these variation rates gives an
estimation of the loss factor $tan(\delta)$.  Our measurement leads to
a loss factor of 0.02, which is four times larger than the loss factor of
GaAs (0.005) \cite{Samara83}.  However, it agrees very well with
values (0.02-0.03) derived form conductance and capacitance
measurements carried out on a similar heterostructure
\cite{Delahaye00c}.
\subsection{Changing the Potential of the Gates}
The frequency dependence of both $\Re e\{Z_{\rm{H}}\}$ and $\Im
m\{Z_{\rm{H}}\}$ has been measured on the LEP2 device for three
different settings of the back-gate potential $V_{\rm{G}}$: at ground, at
twice the Hall voltage and at a potential adjusted to zero the current
coefficient according to the procedure described in \cite{Delahaye01}. 
Figure \ref{RHvsFreq} shows the relative variation of $\Re
e\{Z_{\rm{H}}\}$ with the frequency at three different back-gate
voltages (open symbols).  As expected, the linear frequency
coefficient $\alpha$ decreases when the gate potential is increased.  For
the gate voltage of $2 V_{\rm{H}}$, $\alpha$ becomes negative meaning
that the current $i_{l}$ is now flowing from the back-gate to the
2DEG. The value of the potential leading to a zero frequency
coefficient is slightly above the Hall voltage ($V_{G}=1.3
V_{\rm{H}}$).

The variation of the frequency coefficients, $\alpha$ and
$\beta$, is linear as a function of the back-gate voltage.
The ratio of the slopes of this linear gate voltage dependence gives again
an estimation of the loss factor of the heterostructure. 
Our measurement leads to a value of 0.015 which is consistent 
with the value of 0.02 previously obtained.

\subsection{Removing the Gates}
\begin{figure}%[!b]
\includegraphics[width=\mywdithgraph]{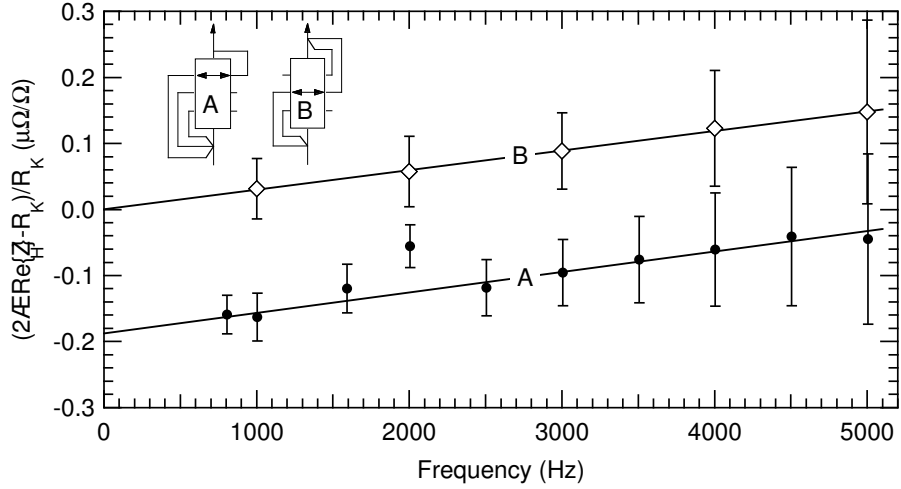}
\caption{Relative deviation of $Re\{Z_{\rm{H}}\}$ from
$R_{\rm{K}}/2$ measured on the ungated LEP3 sample.  The uncertainty
bars correspond to $1~\sigma$.}
\label{RHnoGate}
%\vspace{-0.3cm}
\end{figure}

Figure~\ref{RHnoGate} shows the frequency dependence of $\Re e\{Z_{\rm{H}}\}$
measured on the ungated device LEP3 using an asymmetric multiple-series
connection (A) and a triple-series connection (B).  In both cases, a
small residual linear frequency coefficient of $(0.03\pm 0.03)\cdot 10^{-6}$/kHz
is still observed.  As predicted by our model, the leakage current is
strongly reduced by removing the back-gate. At the moment, it is not clear 
whether this slope comes from a residual leakage current between the 2DEG
and some metallic parts around the sample or from any intrinsic 
properties of the 2DEG, like for example, the 
capacitance associated with the edge states \cite{Jeanneret99}.
An additional difficulty in modelling losses in the 2DEG occurs at 
the corners of the device where the current is injected/extracted 
from the 2DEG. These hot spots where a large potential drop takes 
place over a short distance could also play a role in the residual 
frequency dependence observed in Fig.\ref{RHnoGate}. 

While $\Re e\{Z_{\rm{H}}\}$, measured with the triple-series
connection (B), converges effectively to $R_{\rm{K}}/2$ at zero
frequency, the value obtained with the asymmetric multiple-series
connection (A) differs by $0.2\cdot 10^{-6}$.  We attribute this
deviation to the cable correction term
$(Z/Z_{\rm{H}})^{2}=0.48\cdot10^{-6}$ in eq.~\ref{Zm} for which we use
the room temperature resistance value of the cable (9 $\Omega$). 
Therefore, the calculated correction is slightly overestimated and a
-0.2 $\mu \Omega/\Omega$ offset is observed at zero frequency.
\begin{figure}%[!h]
\includegraphics[width=\mywdithgraph]{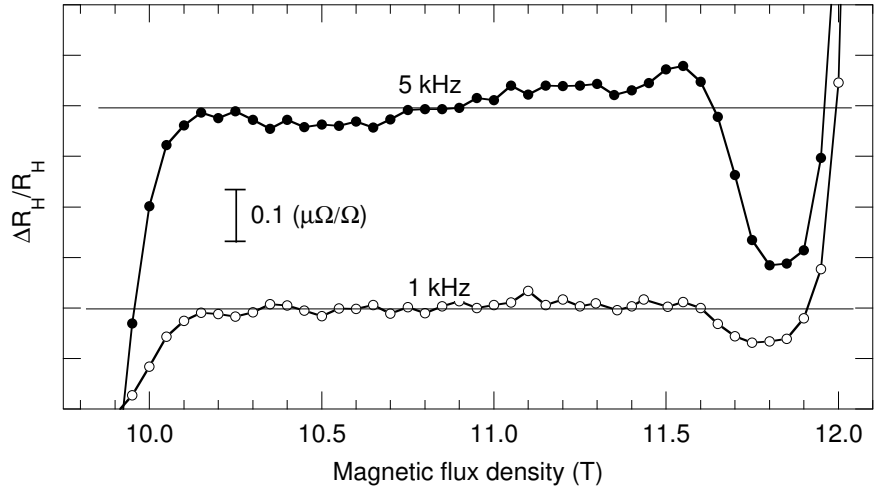}
\caption{Shape of the plateau $i=2$ measured on 
LEP3 with the asymmetric multi-series connection (see configuration A 
in Fig. \ref{RHvsFreq} at 1~kHz and 5~kHz). An 
offset between the two curves has been added for clarity. The 
horizontal lines are a guide to the eye.}
\label{RHvsB}
%\vspace{-0.3cm}
\end{figure}
For the measurements carried out using the asymmetric multiples-series
connection (A), a deviation from the linear frequency dependence is
clearly visible at 2~kHz.  Such a resonance has already been observed
\cite{Chua99} and is attributed to vibrations of the bonding wires. 
Therefore, we excluded the values measured at frequencies 1592~Hz,
2000~Hz and 2500~Hz from the linear fit.  Such a peak does not appear
in the triple-series connection (B) measurements.  Either the
resonance frequency of the bonding wires is out of the frequency range
or the peak amplitude, which is damped by a factor
$(1/R_{\rm{H}})^{3}$, is too small to be observed.

Figure \ref{RHvsB} shows the shape of the plateau $i=2$ measured on
LEP3 in the asymmetric multi-series connection at 1~kHz and 5~kHz. 
The sweep was carried out step by step and each value is the average
of a one minute measurement.  The plateau at 1 kHz is remarkably flat
with a peak-to-peak ripple of only of $0.04\cdot 10^{-6}$ over a field
range larger than 1.4 T. At 5~kHz, the peak-to-peak ripple amounts to
$0.12\cdot 10^{-6}$ over the same field range.  A small feature is present
in the high-B side of the plateau, which is related to a geometrical
effect \cite{vanderWel88}. Indeed, due to the finite width of the voltage
probes, a small component of the longitudinal resistance is mixed into
the Hall resistance.  Otherwise, we did not observe a significant
variation of the plateau width up to the frequency of 5~kHz.

%----------------------------------------------------------------------
% SECTION: Conclusion
%---------------------------------------------------------------------- 
\section{Conclusion}
According to a phenomenological model, the linear frequency dependence
observed with gated quantum Hall samples can be attributed to dielectric losses
produced by leakage currents between the 2DEG and the back-gate.  This
model also explains the effect of the back gate potential on the 
slope of the frequency coefficient.
In a different experiment, the leakage currents were severely
suppressed by simply removing the back-gate.  In this case, a residual
frequency coefficient of $(0.03\pm 0.03)\cdot10^{-6}\rm{/kHz}$ was
obtained without any external potential adjustment.  Here we want to
strongly emphasise that this absence of potential adjustment is
particularly important in view of using the ac quantum Hall effect as
a primary standard of ac resistance.  In addition, the ungated sample
exhibits a broad (around 1.4 T wide) and flat (ripple of 4 parts in
$10^{8}$) plateau making the QHR suitable for high precision ac
metrological applications.  Finally, these results suggest that the
frequency dependence observed in many experiments originates to a
large extent in the measurement apparatus and not in the physical
properties of the 2DEG.

%----------------------------------------------------------------------
% SECTION: Acknolagments
%---------------------------------------------------------------------- 
\section{Acknowledgments}
The authors would like to acknowledge F.~Delahaye for the supply of
LEP devices and H.~B\"{a}rtschi for his skillful assistance.

%\bibliographystyle{IEEEtran}
%\bibliography{Ovbib}

\end{document}